\documentclass[runningheads]{svmult}

\usepackage{makeidx}   

\usepackage{graphicx}  

\usepackage{subeqnar}  

\usepackage{multicol}  

\usepackage{cropmark} 

\usepackage{physprbb}  

\begin{document}

\title*{The Distribution of Dark Matter in Galaxies: the Core Radius  Issue}

\toctitle{Distribution of Dark Matter}


%

%

%

\author{Paolo Salucci \inst{1}}

\authorrunning{Paolo Salucci}

\institute{SISSA, Via Beirut, 2-4
I-34014 Trieste, Ita;y, {\it salucci@sissa.it}}

\maketitle              

\begin{abstract}

I review   the  up-to-date status on the  properties of the Dark Matter density distribution around
Galaxies. The rotation curves of spirals all  conform
  to a same Universal profile which can be uniquely decomposed
as the sum of  an exponential thin stellar disk and a  dark halo  with a
flat density core.
  From dwarfs to giants galaxies, the halos  embedding the stellar component  feature a constant density
region of size $r_0$ and value $\rho_0$, which are  inversely correlated.
The fine structure of dark  halos in the region of  the stellar  disk
has been derived  for   a number of low--luminosity disk galaxies:  the halo circular
velocity  increases almost  linearly with radius out to the edge of the
stellar disk, implying, up there,  an almost
constant dark matter density.  This  sets  a serious discrepancy between
the cuspy density distribution predicted
by N-body simulations of  $\Lambda$CDM  cosmology,  and those actually
detected around galaxies.

The small scatter around the Fundamental Plane (FP) of
elliptical galaxies constraints the distribution  of
 dark  and luminous matter in these systems.
The  measured central velocity dispersion $\sigma_0$ in the FP is  linked to both
 photometric and   dynamical
properties of luminous and dark matter.
As a consequence,  the well-known  features of the FP imply that,  inside  the effective radius $R_e$, the
 stellar spheroid must  dominate over the dark matter,
in contrast with $\Lambda$CDM predictions.

\end{abstract}

\begin{figure}[t]
\begin{center}
\vspace{0.48truecm}
\includegraphics[width=.8\textwidth]{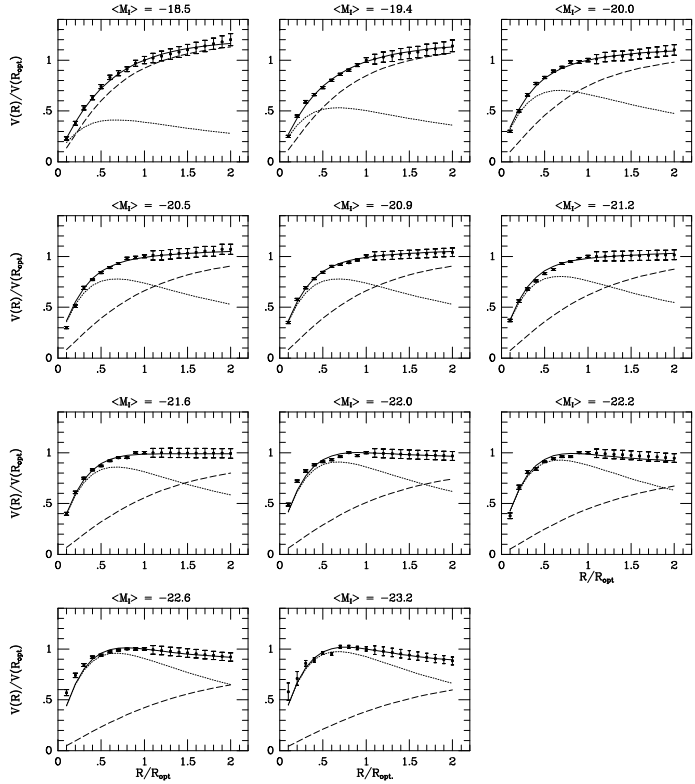}
\vspace {-3truecm}
\end{center}
\caption[]{Synthetic rotation curves (filled circles with error bars) and
the
Universal Rotation Curve (solid line).  The separate dark/luminous
contributions are indicated by a  dotted line (disk)  and a dashed line (halo).}
\end{figure}

\section{Introduction}

Rotation curves (RC's) of disk galaxies are the best probe for dark matter
(DM) on galactic scale. Notwithstanding the impressive amount of knowledge
gathered in the past 20 years, only very  recently we start to shed light on crucial
aspects of the mass  distribution   of  Dark Matter in galaxies, including its
radial density profile, and its Universality.

On a cosmological side, high--resolution  N--body simulations have shown
that cold dark matter (CDM) halos achieve a specific equilibrium density
profile
[13 hereafter NFW, 5, 8, 12, 9] characterized by one free parameter, e.g.
the halo
mass. In the innermost  region,  the DM halo density  shows  an average
profile
which is characterized by a power--law  cusp $\rho \sim r^{-\gamma} $,
with $\gamma =1-1.5$ [13, 12,2].  In detail, CDM halos have:
\begin{equation}
\rho_{\rm NFW}(r) = \frac{\rho_s}{(r/r_s)(1+r/r_s)^2}
\end{equation}
where $r_s$ and $\rho_s$  are respectively the characteristic inner radius
and
density. Let us define $r_{\rm vir}$ as the radius within which the mean
density
is $\Delta_{\rm vir}$ times the mean universal density $\rho_m$ at the halo
formation  redshift, the associated virial mass as  $M_{\rm vir}$ and the
halo velocity as
$V_{\rm vir} \equiv G M_{\rm
vir} / r_{\rm vir}$. In the "concordance"  $\Lambda$CDM scenario:
  $\Omega_m = 0.3$, $\Omega_{\Lambda} =0.7$ and $h=0.7$, so that
$\Delta_{\rm vir} \simeq 340$ at $z \simeq 0$.
Let us set  $c \equiv r_{\rm vir}/r_s$, and from simulations $c \simeq  21 (M_{vir}/10^{11})^{-0.13}$,
 then  in CDM framework  circular velocity
of dark  halos $V_{\rm NFW}(r)$ depends only on their virial masses and
takes the form:
\begin{equation}
V_{\rm NFW}^2(r)= V_{vir}^2 \frac{c}{A(c)} \frac {A(x)}{x}
\end{equation}
where $x \equiv r/r_s$ and $A(x) \equiv \ln (1+x) - x/(1+x)$.
\begin{figure}
\begin{center}
\vspace{-2.7truecm}
\includegraphics[width=.5\textwidth]{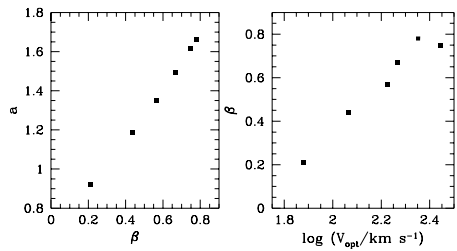}
\end{center}
\caption[]{$a$ {\it vs.} $\beta$ and  $\beta$ {\it vs.} $V_{opt}$.}
\end{figure}

From the  observational point of view, only  recently  the difficulties
in deriving
the  internal structure of halos from available  kinematics  have been
overcome. This has been done
\noindent {\it i)} by means of a careful  study of the  Universal Rotation
Curve [16]  built out of $~ 1000$ individual RC's,
\noindent {\it ii)} by adopting   an halo velocity profile that, out to
$r_{opt}$, is   {\it neutral}  with respect to various different galaxy mass models:
\begin{equation}
V_{h,URC}^2(x)= V^2_{opt}\ (1-\beta)\ (1+a^2)\ {x^2 \over (x^2+a^2)}
\end{equation}
where $x \equiv r/r_{opt}$, $a$ the halo core radius in units of
$r_{opt}$   and $\beta \equiv (V_{d,URC})/V_{opt})^2$ at $R_{opt}$.
Then, by varying
$\beta$  and  $a$, $V_{h,URC}$  can  reproduce the maximum--disk,  the
solid--body, the  no--halo, the all--halo, the NFW
mass models.  (e.g. CDM halos with concentration parameter $c=5$ and
$r_s=r_{opt}$  are well fit by (3) with $a \simeq 0.33$)
\noindent {\it iii)} by means of a number of high-quality high-resolution
individual RC's [1] leading  to trustworthy mass distributions.
Let us define hereafter  $r_d$ as  the
disk scale--length and $V_{opt} \equiv V(r_{opt})$.

\section{Dark Halos Properties from the Universal Rotation Curve}

Let us remind the observational framework: {\it a}) the mass in spirals is
distributed  according to the Inner Baryon
Dominance (IBD) regime [16]: there exists  a characteristic transition
radius
$r_{IBD} \simeq 2  r_d (V_{opt}/220 \ {\rm km/s})^{1.2}$  for which,
at $r \leq r_{IBD}$, the luminous matter completely accounts for the
gravitating mass, whereas,  at $r > r_{IBD}$,  the dark matter shows  up in the kinematics
and {\it
rapidly}  becomes the dominant  mass  component [20, 18, 1].   Then,
although dark
halos  extend down to the galaxy centers,  it is only for $r > r_{IBD}$
that they give  non--negligible contributions to the circular velocity.
{\it
b}) DM is distributed in very  differently way with respect to  the baryons
  [16, 6],  and {\it c}) the  HI contribution to the circular
velocity, at $r < r_{opt}$, is small [e.g. 17].

Persic, Salucci and Stel [16] have derived  from  $\sim 20000$ velocity
measurements, relative to  $\sim 900$ rotation curves,   $V_{syn} ({r \over {r_{opt}} }; M_I)$,
 the synthetic rotation velocities of spirals binned in  intervals of magnitudes.
Each individual RC's (see Fig. (1))  shows  a  variance, with respect to  synthetic
curves of the corresponding magnitude,  smaller than observational
errors: spirals sweep a very narrow locus in the RC-~profile/amplitude/
luminosity space. Thus, as regard the average main properties of the DM
distribution, eq (3) is equivalent to a  large sample of individual objects.

The whole set of synthetic  RC's define the Universal Rotation Curve (URC)
that we represent analytically with the sum of two terms: {\it a}) the standard
exponential thin disk term:
\begin{equation}
V^2_{d,URC}(x)=1.28~\beta\ V^2_{opt}~ x^2~(I_0K_0-I_1K_1)|_{1.6x}
\end{equation}
and the  spherical halo term given by (3). The data (i.e. the synthetic
curves $V_{syn}$) select the actual mass model:
by setting  $V_{URC}^2(x)=V^2_{h,URC}(x,\beta,a)+ V^2_{d, URC}(x,\beta) $
with $a$ and $\beta$ as free parameters, an extremely  good fit occurs when:
$ \beta=\beta (\log V_{opt})$  and  $a =a(\beta)$ as plotted  in
Fig. (2):
the URC  reproduces  $V_{syn}(r)$ up to its {\it rms} (i.e.  within
$2\%$).  Moreover,  at a fixed luminosity,   the $1\sigma$ fitting
uncertainties for $a$ and $\beta$ are lesser than   20\%.

The emerging scenario is the following: inside $r_{\rm opt}$ smaller objects have
larger dark-to-stellar  matter :  $M_{\ast}/M_{\rm vir} \simeq 0.2\
(M_{\ast}/2 \times 10^{11} M_{\odot})^{j}$ ($j \sim 0.75$)  [20]) and within {\it each}  galaxy
the  dark mass increases
with radius with a power-law  exponent between 2 and 3.

\begin{figure}[t]
\vskip -2.1truecm
\includegraphics[width=1.\textwidth]{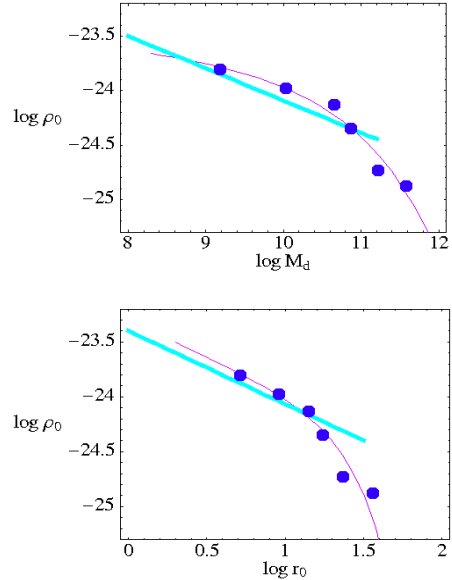}
\vskip -7.3truecm
\caption[]{ {\it up)} Central halo density $\rho_0$  (in ${\rm g/cm}^3$)
{\it vs.}  disk mass (in solar units) for normal spirals ({\it filled
circles}); {\it bottom)} central
density {\it vs.} core radii (in kpc) for normal  spirals (see [19]).}

\end{figure}

This  evidence calls for a cored  Dark Matter density [3, 4, 1]. Then,  we are allowed
to  pass from  the "neutral" distribution
of DM of eq (3) in which a  "core" may appear in the {\it velocity}  profile (i.e. in a quantity which is  directly measured),  to the
much more specific mass distribution  given by the Burkert {\it density} profile  that forces a core radius into  the NFW profile.
\begin{equation}
\rho_{\rm B}(r) = \frac{\rho_0\ r^3_0}{(r+r_0)(r^2+r_0^2)}
\end{equation}
with  $\rho_0$ and $r_0$  free parameters (the central
DM density and the core radius). Then
$M_{\rm B}(r) = 4\ M_0\ \{ \ln (1 + r/r_0)  -\arctan (r/r_0) + 0.5  \ln
[1+(r/r_0)^2]\}$
with $M_0 \simeq 1.6\ \rho_0\ r_0^3  $ the dark  mass within $r_0$.
The halo contribution to the circular velocity is: $V^2_{\rm h,B }(r) =G\ M_{\rm B}(r)/ r$.

The Disk + Burkert halo  model  applied to the synthetic rotation curves  leads to
core parameters $r_0$, $\rho_0$ strong  correlated and linked to the disk
mass : dark halos  behave as  an 1--parameter family, completely specified by
{ \it e.g.}  their central density
$\rho_0$ (see Fig. (3) ).

These relationships imply that the densest halos harbor the least massive
disks (see Fig. 3) while  the feature
of the  curvature at the highest
masses/lowest densities  in the  $\rho_0$ {\it vs.} $r_0$ relationship may be  related
  to  the existence of an {\it upper mass limit} in $M_{vir}$ at $few  \times  10^{12} M_\odot$.

\subsection{Testing CDM with the URC}

From the analysis of the URC one concludes that
dark halos are not kinematically cold structures,  but
``warm" regions of  sizes $ r_0 \propto \rho_0^{-1.5}$
with $r_0 \sim 4-7\ r_d$. Then, the boundary of the core region
is beyond the region  of the stellar disk. There is no
evidence that the DM density
converges to a $ r^{-1}$ (or a steeper) regime, as dictated by CDM scenario.
\begin{figure}
\vskip -1.7truecm
\begin{center}
\includegraphics[width=1.1\textwidth]{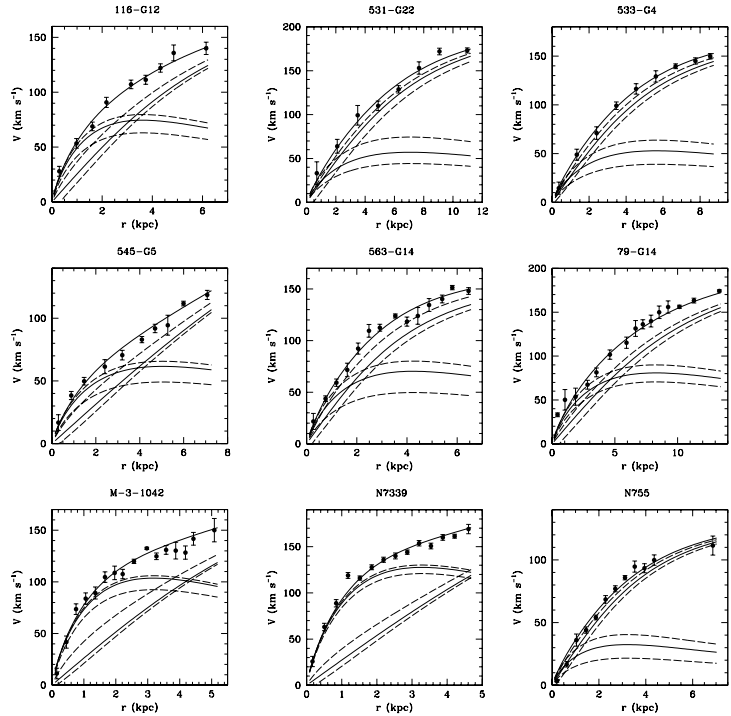}
\vskip -8.2truecm
\end{center}
\caption[]{ B+ Disk  models ({\it thick solid line})   ({\it points with
errorbars}).  Thin solid lines represent the   disk and halo contributions.
The maximum and minimum  disk
solutions  ({\it dashed lines}) provide the  uncertainties see [1]}
\end{figure}
\section{Dark Matter Properties from Individual Rotation Curves }

Although deriving halo densities from individual RC's is certainly
complicated,
the belief according to which  one is bound to get ambiguous  halo  mass
modeling
[as claimed in some  work]
 is not always correct. In fact, this is true only for rotation
curves of low spatial resolution,
i.e. those  with  less than $\sim3$ measures per exponential disk length--scale,
as   most of HI rotation curves.  In this case, since the parameters of the galaxy
structure
are crucially sensitive to the {\it shape} of the rotation curve in the region
$0<r<r_d$,
there are no sufficient data to constrain the mass model. In the case of
high--quality {\it optical}
RC's  tens of independent measurements solve the problem.
Moreover, since the dark component can be better traced when the disk
contributes
to the dynamics in a modest way, a convenient strategy leads  to investigate
DM--dominated
objects, like dwarf and low surface brightness (LSB) galaxies.  For
the latter   [e.g. 7, 11, 3, 4, 9, 10, 21, 23] the results are not
definitive in that they unfortunately are intrinsically
uncertain, due to the limited quality
of their  kinematics.

Since the observed RC's  have  Universal  features and  most of the
properties of cosmological halos are
also  claimed universal,  an
useful strategy is to investigate  a restricted  number of high--quality
{\it
optical} rotation curves  of {\it low luminosity } late--type spirals, with
$I$--band absolute magnitudes $-21.4<M_I<-20.0$  and  $100 <V_{opt}< 170$ km
s$^{-1}$. Objects in this luminosity/velocity range are DM dominated [e.g.
20]
but  their optical  RC's, measured at a typical angular resolution of $2^{\prime
\prime}$,
have the  excellent  spatial resolution of  $\sim 100 (D/10 \ {\rm Mpc})$
pc and $n_{data}
\sim r_{opt}/w$ independent measurements.  For nearby galaxies: $ w<< r_{d}$
and $n_{data}>25$.

In [1] we extracted, from the `excellent' subsample  of $80$ rotation
curves of [15],
the best 9 rotation curves. These RC's  (as any  rotation curve that candidates itself to yield
something crucial on the DM distribution)  trace properly
the gravitational potential since satisfy the following  quality requirements: {\it 1)}  data extend at least out to
the optical radius, {\it 2)} they
are smooth and  symmetric, {\it 3)} they have small internal {\it rms}, {\it 4)} they
have high  spatial resolution and a homogeneous radial data coverage of
$30-100$ data points  between
the  two arms. Each RC has $7-15$
velocity points inside  $r_{opt}$, each one being the average of $2-6$
independent data.
The RC's  spatial resolution is better than $1/20\ r_{opt}$, the velocity
{\it
rms}  is about $3\%$ and  the RC's logarithmic derivative is generally known
within  about 0.05.

We model the mass distribution as the sum of two components: a stellar disk
and a spherical dark halo, therefore:
$V^2(observed)=V^2(disk)+V^2_h(halo)$.
  Light traces the stellar  mass via the radially constant
mass--to--light ratio.
We neglect the gas contribution
$V_{gas}(r)$ since   in normal spirals it is small [17,  Fig. 4.13]:
$ \beta_{gas}\equiv (V^2_{\rm gas}/V^2)_{r_{opt}}
\sim 0.1$.    Incidentally, this is not
the case for dwarfs and LSB´s: most  of their kinematics is affected in
different ways by the HI
disk gravitational pull.  The  disk contribution to circular velocity
is given by (4),  while the dark  halo contribution  by (3).  Finally we
normalize (at  $r_{opt}$) the velocity model $(V_d^2+V^2_h)^{1/2}$ to the
observed rotation  speed $V_{opt}$.

For each galaxy, we determine the values of the parameters $\beta $ and $a$
by means of a $\chi ^2$--minimization fit to the observed rotation curves:
\begin{equation}
V^2_{model}(r; \beta , a) = V^2_d (r; \beta) + V^2_h (r; \beta , a)
\end{equation}
A central role in performing the  mass decomposition is
played by the derivative of the velocity field $dV/dr$. It has been shown
[e.g. 14] that by taking into   account the logarithmic gradient of the
circular velocity field defined as:  $\nabla (r)\equiv \frac{d \log V(r)}{d
\log r} $
one best  retrieves the  crucial  information  stored in the shape of
the rotation curve.   Then,  we set the   $\chi^2$ statistics as the sum of
$\chi^2$'s
evaluated on velocities and  on logarithmic  gradients. In detail, by setting
$ \chi^2_V =\sum^{n_V}_{i=1}\frac{V_i-V_{model}(r_i; \beta,a)}  {\delta V_i}$
and  $\chi^2_{\nabla} =
\sum^{n_{\nabla}}_{i=1}\frac{\nabla(r_i)-\nabla_{model}
(r_i; \beta,a)}{\delta \nabla_i}$.
we minimize  the
quantity
\begin{equation}
\chi^2_{tot} \equiv \chi^2_V+ \chi^2_{\nabla}
\end{equation}
to get the mass model. 

Let me point out that any claim of "different  mass models" that all  would
account for  a certain   rotation curve, does instead originate from the  (low)  quality of the latter  that does not allow
a reasonably accurate derivation  of  $\nabla(r)$.

The best--fit models parameters for the "neutral" distribution of (3) are shown in Fig. 4. The
disk--contribution $\beta $ and the  halo core radius $a$ span a range from
0.1 to 0.5 and from 0.8 to 2.5, respectively. They are pretty well
constrained in
a small and continuous region of the ($a$, $\beta$)  space.
It is obvious that the halo curves
are increasing almost linearly,  out to the last data point.
Remarkably, we find that the size of the halo density core is always greater
than the disk characteristic scale--length $r_d$ and it can extend beyond
the
disk edge (and the region investigated).

\begin{figure}[t]
\vspace{-1.6truecm}
\begin{center}
\includegraphics[width=12.5truecm, height=17truecm]{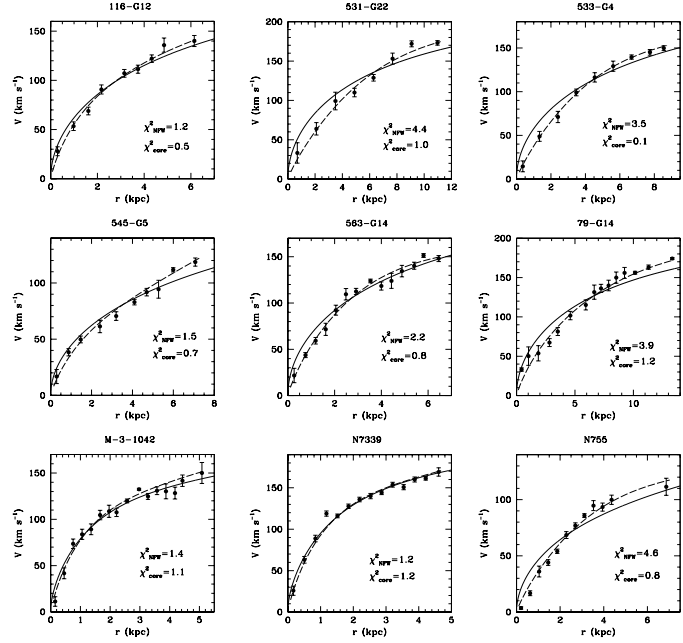}
\end{center}
\vspace{-6.5truecm}
\caption[]{NFW best--fits {\it solid lines} of the rotation curves  {\it
(filled circles)} compared with   the URC halo  + Disk   fits {\it (dashed lines)}. The
$\chi^2$ values are also indicated.}
\end{figure}

\subsection{Testing CDM}

 Let us assume for the dark halos the  NFW functional form  given by (1) and  fit the 9 RC's
leaving $c$ and $r_s$ as free independent parameters,
although N--body simulations and semi-analytic investigations indicate  that
they correlate in order  to increase the chance of a good  fit. It must be assumed, however, for these 9 test objects,
a conservative halo mass upper limit of
$2 \times 10^{12} M_\odot$.
The fits to the data are shown in  Figs. (5) and (6), together with the URC fits: for
seven out of  nine   objects the NFW models are unacceptably worse than the
URC  solutions.   Moreover,  the resulting CDM  disk  $I$-band
mass--to--light ratios  turn out to be in some cases  $ \sim 0.01  $ solar units,  i.e.  unacceptably
low  in the $I$-band.

 We definitely  conclude that  there is no shortage of dark halos around objects with a  trustable rotation curve,
  that show a density
distribution inconsistent with that predicted by  collision-less CDM.

\begin{figure}[t]
\vspace{-1.2truecm}
 \begin{center}
\includegraphics[width=0.8\textwidth]{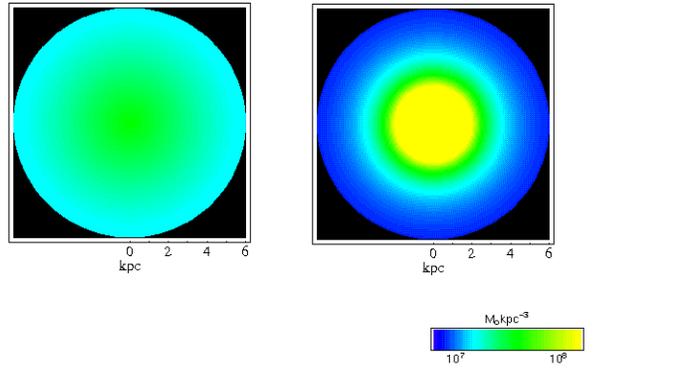}
\end{center}
\vspace{-7.7truecm}
\caption[]{ {\it left})  The density of the  dark halo of   116--G12, {\it right})  The CDM prediction }
\label{eps2}
\end{figure}

\section{ Halos around Ellipticals}

The very low scatter that  ellipticals show lying
on the Fundamental Plane is a  kinematical
feature can be  used to investigate their  Dark Matter distribution.
The central velocity dispersion $\sigma_0$ is the key quantity,  in that it
strongly depends on the mass distribution of both stellar  and dark matter
[34].
A  sample of  221 E/S0 galaxies with  good photometric
and spectroscopic data has been used to investigate this point.  This  sample defines the standard
  "empirical" FP in the
coordinate space  of ($log \sigma_0$, $log R_e$, $log L_r$). The galaxies
have a average distance of
0.084 dex with respect to the plane,  of the order of the
measurement uncertainties, estimated as large as
$\pm 0.05$ dex.

It is easy to show, for the  NFW distribution of $\Lambda$CDM cosmology, that  $\sigma_0$ is a
specific function of
the total virial mass and the stellar mass-to-light ratio, with no other
free parameters.
For the Burkert (1995) density distribution,  $\sigma_0$ depends instead on
the stellar mass-to-light ratio,
the value of the core radius $r_0$ and the fraction of dark mass inside  $R_e$.
\begin{figure}[t]
\begin{center}
\hspace{-3.5truecm}
\includegraphics[width=15truecm]{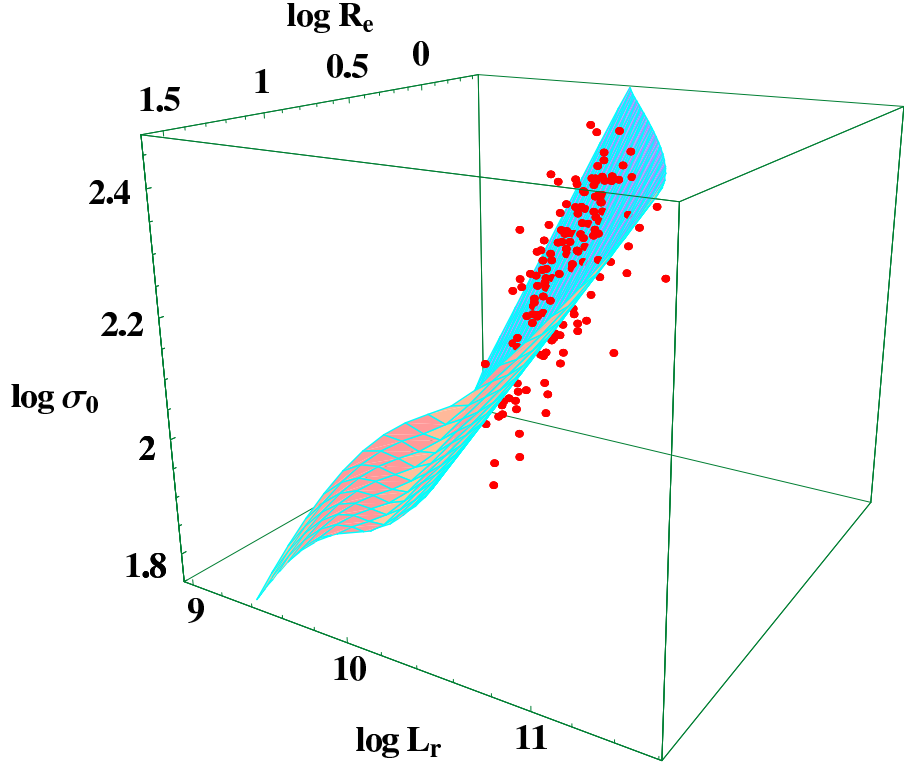}
\end{center}
\vspace{-17.7truecm}
\caption{The  CDM surface  in the log--space ($R_e$, $L_r$, $\sigma_0$) compared to
sample data (points). $R_e$ in kpc, $L_r$ in $L_{r \odot}$
and $\sigma_0$ in km/s.}
\label{FPN}
\end{figure}

It is shown that the  existence of a tight FP relating the above quantities implies that ellipticals
are largely dominated, within $R_e$, by the stellar spheroid,  independently
of the actual  DM distribution. However,  $\Lambda$CDM predicts large amounts of dark matter inside $R_e$
in view of the cuspy density  distribution of CDM halos: as results  in this framework the
ellipticals would lie
on a {\ it curved surface},  inconsistent with the observed {\it
plane} (see Fig. (7)).

\begin{figure}
\begin{center}
\hspace{-3.5truecm}
\includegraphics[width=15truecm]{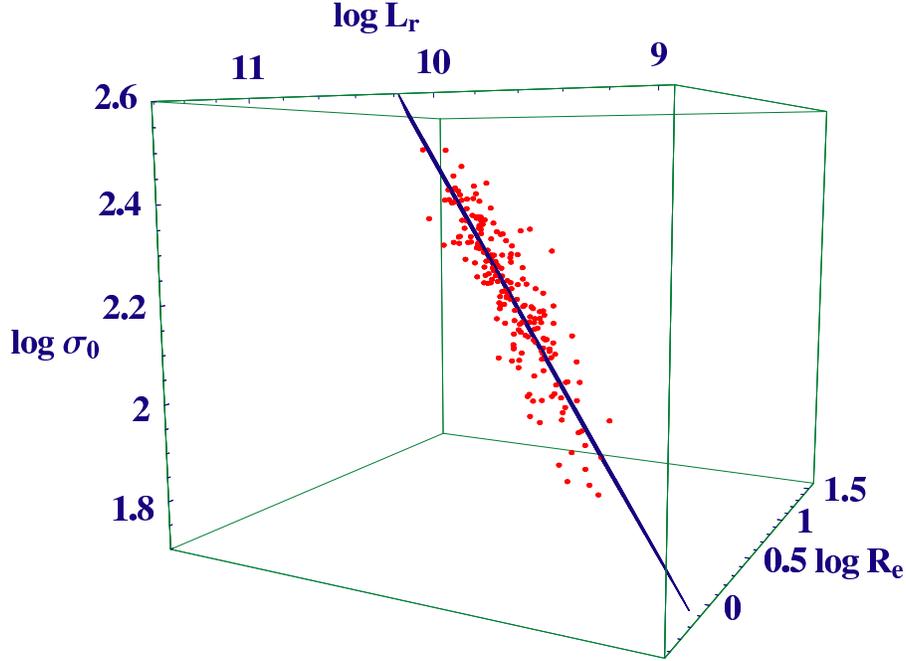}
\end{center}
\vspace{-19.2truecm}
\caption{The plane from  Spheroid+Burkert halo  model. }
\label{FPB}
\end{figure}

The  Burkert  density distribution, in which substantial amounts  dark matter  can be placed outside $R_e$,
 leads to a relationship that is a plane
resulting  in perfect  agreement
with the observed one (see Fig (8)). This  implies a dark--to--luminous mass fraction
within the effective
radius of about $0.3\pm 0.2$ and a luminosity
dependence  of the spheroidal mass--to--light ratio:
$M_{sph}/L_r=(5.3\pm 0.1) (L_r/L_{\ast r}) ^{0.21\pm 0.03}$,
  in Gunn--$r$ band. Moreover, as a firm constraint, we can state  $r_0>R_e$.
\begin{figure}[t]
\begin{center}
\hspace {-0.0truecm}
\includegraphics[width=10truecm, height= 10truecm]{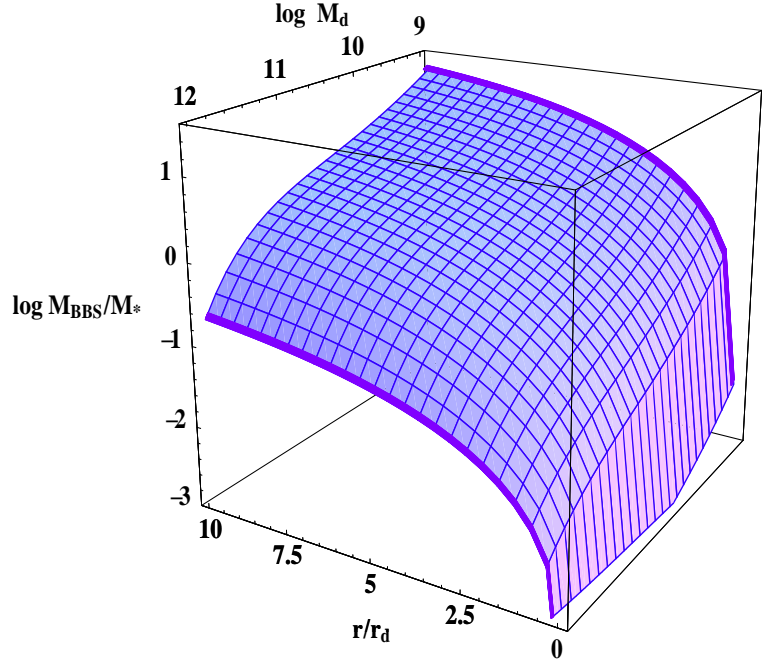}
\end{center}
\vspace {-1.0truecm}
\caption[]{The dark--to--luminous mass ratio as function of the normalized
radius and the total disk mass. BBS halo  is the Burkert profile  as in Borriello and Salucci [1]}
\end{figure}

\section {More  Support for  Core Radii. }

The Trieste group has provided a crucial evidence on the  "core radii" issue,  however,  results from  other
investigations are also very important and must be considered,  in that, alongside with those  referred
in the previous sections, build a formidable case for the existence of constant density cores at the 
centers of the dark halos surrounding  spiral galaxies.

\begin{table}
\caption{Extra Evidence }
\label{tab_param_M82}
\begin{center}
\begin{tabular}{ll}
\hline
\hline
Author(s)           & Paper \\
\hline
Ortwin et al   &   AJ  121, 1936 \\
Fuchs  &  astro-ph/0212485 \\
Swaters et al . &  ApJ 583, 732 \\
 Weldrake et al.  &  MN  340, 12 \\
 Bottema \&   & Verheijen A\&A  388, 793 \\
 Bolatto et al.  & ApJ  565, 238 \\
 de Blok et al. &  AJ 122,  2396 \\
de Blok et al.  &  ApJL 552,  23 \\
Stil \& Israel  &   A\&A  392,  473 \\
 Dutton et al. &   astro-ph/0310001 \\
Marchesini et al. &    ApJ. 575,  801 \\
Fraternali et al. &  AJ 123, 3124 \\
\hline
\end{tabular}
\end{center}
\end{table}

\section{The Intriguing Evidence from Dark Matter Halos}

\begin{figure}[t]
\vspace {-1.99truecm}
\begin{center}
\includegraphics[width=.9\textwidth]{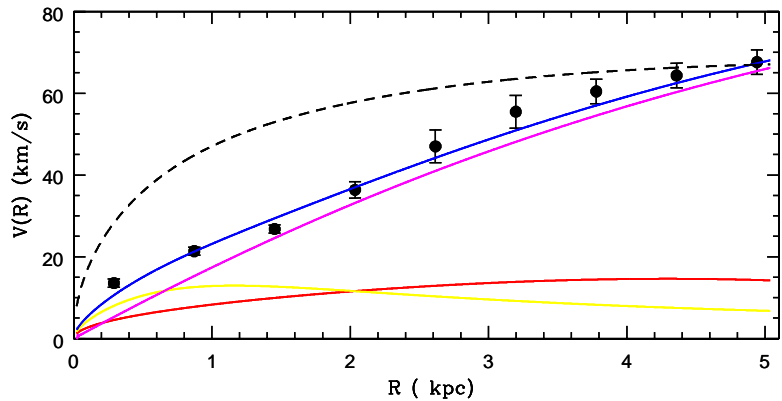}
\end{center}
\vspace{-3.3truecm}
\caption{DDO 47 rotation curve vs different mass models. The continous line is the
sum of the stellar (red),
the gas  (yellow) and the Burkert halo distributions (magenta). The  dashed line represents is the NFW
distribution. }
\end{figure}

From observations the dark halos around galaxies  emerge as an
one--parameter family;
the order parameter (either the central density or the core
radius)  correlates with the stellar mass. However,   the global structural properties of the dark halo, like the
virial radius or the virial mass
could be  extrapolated only  empirically,
because we  do not know how to prolongate, outside the region
mapped by data and  in a theoretical way,
the DM profile. In fact,  inside the stellar disk region,  the halo RCs are
determined by physical parameters, the central core density and the core
radius,  that have no  counterpart in the gravitational
instability/hierarchical clustering picture.

Important  relationships
among physical  quantities still emerge  from the mass modelling:
in Fig. 9 we show the dark--to--luminous mass
ratio as function of the normalized  radius and of  the total disk mass. The
surface has been obtained by adopting the correlations between the halo and
the disk parameters derived in  [19].
  Therefore the   dark--to--stellar  mass ratio, at fixed fraction of disk length-scales,
increases as  the total disk mass
decreases;  for example: in the range  $ 1<r/ r_d \leq 3 $ it raises by  $20\%$ for massive
disks ($M_d=10^{12} M_{\odot}$) while it raises by  $220\%$  for  smaller disks ($M_d=10^9 M_{\odot}$).

Two conclusive statements can be drawn:   dark matter halos
have an inner  constant--density region, whose size exceeds the stellar
disk length--scale. Second, there is no evidence that dark halos converge,
at
large radii, to a $\rho \sim r^{-2}$ (or steeper) profile.

The existence of a region of  ``constant"  density    as wide as the stellar disk   is hardly
explained  within current theories of galaxy formation. A number of different
solutions have been proposed to solve this problem [e.g. 30, 31, 32, 33].
Before commenting on them let us stress that any solution of the "core radius" issue must
account for  all the intriguing
halo properties described in this review. Let us point out that we  review the several tenths of
 attempts for a solution  to
the "core radii" issue will be reviewed elsewhere. He we  just classify them in families:

1- Dark Matter "interacted" with itself or with baryons. Original "cuspy" halos have been smoothed in this way.

2- Dark Matter has a different power spectrum/perturbations evolution than the current Standard Picture
and this has   "produced"  the cores.

3- Dark Matter is a "field", that  naturally mimicks the effects of a cored halo of particles.

4- Within the standard  $\Lambda$CDM scenario,
N-Body simulations have failed,  for a  variety of reasons,  to discover this  intrinsic and   real  feature.

\section{Prologue}

 Let me  finish where I  started: i.e.   by showing a test case for the existence of a core in the density distribution of
the dark halo around  a galaxy. DDO 47, see Fig (10)  has a rotation curve that increases linearly from the first
data point, at $300 \ pc$, up to the last one, at $ 5 \ kpc$ well beyond the stellar  disk edge.
The  RC  profile implies the presence
of a dominating (dark) halo with an (approximately) constant density out to the last measured point,
 {\it prior}  any mass modelling.

\section{Acknowledgments}

P.S thanks G. Danese for stimulating discussions and  I. Yegorova for help in the presentation.

\end{document}